# Characterizing oxygen atoms in perovskite and pyrochlore oxides using ADF-STEM at a resolution of a few tens of picometers

Ali Mostaed[1], Brant Walkley[2,3], Monica Ciomaga Hatnean[4], Geetha Balakrishnan[4], Martin R. Lees[4], Richard Beanland[4], Derek C. Sinclair[2] and Ian M. Reaney[2]

[1] *Department of Materials, University of Oxford, Parks Road, Oxford OX1 3PH, UK*
[2] *Department of Materials Science and Engineering, University of Sheffield, Sheffield S1 3JD, UK*
[3] *Department of Chemical and Biological Engineering, University of Sheffield, Sheffield S1 3JD, UK*
[4] *Department of Physics, University of Warwick, Gibbet Hill Road, Coventry CV4 7AL, UK*

**Abstract**

We present an aberration corrected scanning transmission electron microscopy (ac-STEM) analysis of the perovskite ($LaFeO_3$) and pyrochlore ($Yb_2Ti_2O_7$ and $Pr_2Zr_2O_7$) oxides and demonstrate that both the shape and contrast of visible atomic columns in annular dark-field (ADF) images are sensitive to the presence of nearby atoms of low atomic number (e.g. oxygen). We show that point defects (e.g. oxygen vacancies), which are invisible – or difficult to observe due to limited sensitivity – in x-ray and neutron diffraction measurements, are the origin of the complex magnetic ground state of pyrochlore oxides. In addition, we present, for the first time, a method by which light atoms can be resolved in the quantitative ADF-STEM images. Using this method, we resolved oxygen atoms in perovskite and pyrochlore oxides.

1. Introduction

Ceramic oxides with perovskite ($ABO_3$) or pyrochlore ($A_2B_2O_7$) structure, where *A* and *B* are cations with very different sizes, exhibit interesting physical and electrical phenomena including magnetic [1, 2], ferroelectric [3-6], dielectric [7-9], pyroelectric [4, 10], piezoelectric [4, 11] and superconducting [12, 13] properties. The functionality of these oxides is often dependent on microstructural defects, in particular oxygen displacement or oxygen vacancies [14-19]. For example, local structural defects (e.g. vacancies) in the oxygen lattice can induce small changes in the $Pr^{3+}$ ion network, and thus affect the magnetic properties of praseodymium zirconate ($Pr_2Zr_2O_7$). It has been demonstrated in x-ray diffraction and neutron scattering measurements that off-centring of $Pr^{3+}$ ions from the ideal crystallographic position, which can be due to presence of oxygen vacancies, is at the origin of the complex magnetic ground state of single-crystal $Pr_2Zr_2O_7$ [20-23]. However, oxygen vacancies themselves remain undetected [20-23]. Although x-ray and neutron diffraction techniques have been widely used to study the presence and position of oxygen in oxide materials [24-28], these techniques only provide information regarding the average structure, as the information is collected from a large volume of a specimen. Transmission electron microscopy (TEM), on the other hand, offers the opportunity to probe their local structure. For instance, aberration corrected scanning transmission electron microscopy (ac-STEM) can be exploited to obtain information about the structure of materials with sub-angstrom resolution. Nevertheless, detecting oxygen, which is a light element, in the presence of heavier elements in oxides remains challenging.

Bright-field (BF) and annular bright-field (ABF) imaging techniques have been widely used to reveal the position of oxygen in oxide materials [29-34]; however, as intensities in ABF images are not related in a straightforward way to atomic number, they are not reliable to study the oxygen occupancy. On the other hand, annular dark-field (ADF) imaging provides quantitative information regarding the atomic structure of materials as the contrast of visible atom columns in ADF images is proportional to the atomic number ($Z$) of the atoms in those columns, i.e. ADF is a $Z$-contrast imaging technique [34-37]. The contribution of each single atom ($Z^n$; where *n* is a constant with a value between 1.5 to 2 [38, 39]) in the specimen to the ADF image is proportional to the atomic number of that atom as well as the electron probe intensity at that atom, $J(r)$, which can be simply written as a differential contribution ($dI$) as follows:

$$dI(\boldsymbol{r_p}, z) = Z^n J(\boldsymbol{r_a} - \boldsymbol{r_p}, z) \,, \tag{1}$$

where, $\boldsymbol{r_p}$ is the probe position, $\boldsymbol{r_a}$ and $z$ are the lateral position and depth of the atom [40]. When the probe is at or very close to an atom column containing heavy atoms, there would be a sufficient attractive Coulombic force from the heavy atoms to trap or focus the majority of the electron beam along the atom column and as a consequence avoid the probe spreading out from the column [41]. Thus, the contribution of heavy atom columns to the ADF intensity is significant, as not only do they contain heavy atoms (high $Z^n$ in Eq. 1) but also the majority of the probe intensity distribution is on-column (high $J(r)$ in Eq. 1). On the other hand, as the attractive Coulombic potentials for the atom columns containing light atoms are too weak to trap the electron beam, probe de-channelling can occur when the probe is at or very close to these columns and, therefore, a fraction of the probe intensity can spread on their nearby atom columns, i.e. low $J(r)$ [40, 41]. Thus, columns with light atoms contribute far less than columns with heavy atoms to the intensity of ADF images as both their $Z^n$ and $J(r)$ values are significantly smaller for columns with the lighter atoms. This is the reason why light elements, especially in the presence of relatively heavier elements, e.g. oxygen atoms in oxides, are invisible in ADF images. In contrast to the aforementioned reasons for invisible oxygen atoms in ADF images, Mostaed et al. [15] showed that the detailed intensity distribution around the visible cation columns in ADF-STEM images obtained from the pyrochlore $Yb_2Ti_2O_7$ is sensitive to the presence of their nearby invisible light atoms. In the current work, we explain how light atoms like oxygen can contribute sufficiently to the intensity of ADF images to be detected.

Here we use ac-STEM to study the atomic structure of pyrochlore ($Yb_2Ti_2O_7$ and $Pr_2Zr_2O_7$) and perovskite ($LaFeO_3$) oxides and demonstrate that the intensity as well as shape of visible cation columns in ADF-STEM images are sensitive to the presence of their nearby invisible oxygen atoms. We propose a novel method by which the invisible oxygen columns can be resolved in the ADF-STEM images. Using this method, we, for the first time, resolve the position of oxygen columns in ADF-STEM images obtained from pyrochlore ($Yb_2Ti_2O_7$) as well as perovskite ($LaFeO_3$) oxides.

## 2. Experimental methods

$La_{1-x}FeO_{3-\delta}$ ceramics were prepared via a solid-state reaction method using solid precursors $La_2O_3$ (Sigma Aldrich 99.99 wt.%) and $Fe_2O_3$ (Sigma Aldrich 99 wt.%). Stoichiometric amounts of the solid precursors were weighed and mixed by ball milling in isopropanol with yttria-stablised zirconia (YSZ) media for 12h. The mixed powders were dried at 80 ºC overnight and subsequently sieved and calcined in air at 1100 ºC for 6 hours (5 ºC/min ramp on heating and cooling) in an alumina crucible. Calcined powders were then sieved and ball milled in isopropanol with YSZ media for 12h, and subsequently dried at 80 ºC for 12h. Pellets were then prepared from the dried calcined powders using a uniaxial press, and subsequently sintered in air for 6 hours at 1200 ºC (5 ºC ramp rate on heating and cooling in each case) in an alumina crucible.

$Pr_2Zr_2O_7$ Samples were prepared in polycrystalline form by the conventional solid-state method. Powders of $Pr_6O_{11}$ (99.9 %) and $ZrO_2$ (99 %) were weighed in stoichiometric amounts, mixed together and heat treated in air for several days (in four steps) at 1400-1450 °C, with intermediate grindings, see Ref. [42] for more information. The resulting sample, i.e. sample A, was brown in colour. A second praseodymium zirconate sample, i.e. sample B, was prepared under similar conditions, with an additional annealing step where the sample was heat treated in a reducing atmosphere (Ar + 3% $H_2$) at 1200 °C for 2 days. The resulting powder was bright green coloured. Sample *A* is expected to be of the nominal composition $Pr_2Zr_2O_7$, while the reduced sample, *B*, is expected to be of composition $Pr_2Zr_2O_{7-\delta}$.

A doubly-corrected JEOL ARM200F microscope, operating at 200 keV, was employed to study the atomic structure of samples. ADF-STEM images were obtained from flat, defect-free and uniform regions of interest of the samples using an ADF detector where the inner angle of the ADF detector was ~4.6α (α, the convergence semi-angle of the electron probe, was ~15 or ~30 mrad). Up to sixty ADF images, all with a dwell time of 10 μs/pixel, were collected sequentially from a region of interest and subsequentially the images were aligned using both normalized cross-correlation (using in-house code) and non-rigid image registration (using Smart Align software [43]) and then summed to obtain high quality ADF-STEM images with a good (> 30) signal to noise ratio (SNR). The DetectColumns [15] program was used to calculate the position and intensity of atom columns in the atomic resolution ADF-STEM images. Multi-slice frozen phonon (>30 frozen phonons) image simulations were performed with Prismatic [44, 45] and Dr. Probe [46] software to compare with experimental data (see Section I in Supplemental Material for more information).

## 3. Results

### *3.1 Annular dark-field STEM*

Figs. 1(a) shows a representative ADF-STEM image from sample A ($Pr_2Zr_2O_7$) obtained along the [2 1 1] direction. As shown in Fig. 1(b), there are three types of cation columns in this projection; those containing only *A*-site cations (i.e. Pr), those containing only *B* site cations (i.e. Zr), and mixed (*M*) columns containing 50% *A* and 50% *B*. The *M* columns can be divided into two different types, *M1* and *M2*, according to the number and position of their nearby oxygen columns. There are twice as many oxygen atoms around *M1* columns in comparison to *M2* columns. The oxygen atoms around *M2* columns are closer to the centre of the cation columns. As a result, in the ADF-STEM image the *M2* columns have a higher peak intensity while *M1* columns have a higher radial intensity at larger radii (> 70 pm) as displayed in Fig. 1(c). These results, consistent with mean radial intensity profiles obtained from another pyrochlore $Yb_2Ti_2O_7$ [15] (Figs. 1(d)), indicate that invisible light atoms like oxygen have an influence on the contrast of their nearby visible cation columns with a relatively high atomic number in ADF images.

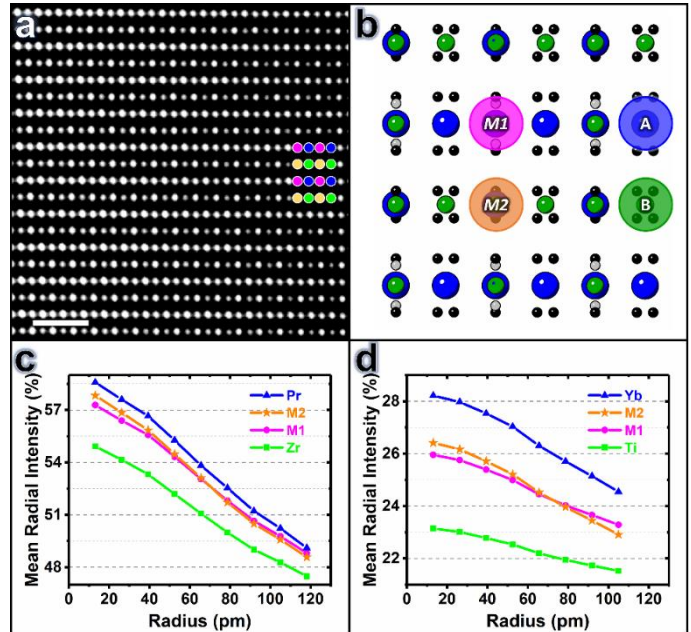

**Fig. 1.** (a) ADF-STEM image taken with an ADF detector inner angle of 2.4α from sample A ($Pr_2Zr_2O_7$) viewed along [2 1 1] (scale bar is 1 nm). The four visible atom columns in this image, i.e. Pr, Zr, *M1* and *M2* columns, are shown in blue, green, pink and orange at the right of the image. (b) [2 1 1] projection of the crystal structure of pyrochlore $A_2B_2O_7$ showing the four different types of atom columns. Here, *A*, *B*, O(*48f*) and O(*8b*) atoms are shown in blue, green, black and grey, respectively. (c) Mean radial intensity profiles for the four atom columns, i.e. Pr, Zr, *M1* and *M2*, in (a) calculated using DetectColumns [15] program. (d) Mean radial intensity profiles for Yb, Ti, *M1* and *M2* atom columns in the ADF image of pyrochlore $Yb_2Ti_2O_7$ (Sample 1 in Ref. [15]) adapted from Ref. [15].

As illustrated in Fig. 2, we also investigated the intensity distribution around visible atom columns in ADF images obtained from sample B, which is expected to be oxygen deficient ($Pr_2Zr_2O_{7-\delta}$). Interestingly, here we observe that the radial intensity of *M2* columns is always more than that of *M1*, in other words, the radial intensity of *M1* never exceeds that of *M2*. This could be due to the presence of oxygen vacancies, as Mostaed et al. [15] have already reported that oxygen vacancies alter the radial intensity distribution around *M* columns in ytterbium titanate. However, in the case of oxygen-deficient ytterbium titanate ($Yb_2Ti_2O_{7-\gamma}$), the intensity of *M1* columns is always greater than that of *M2* (Fig. 2(c)) while the intensity of *M2* columns is always greater than that of *M1* in $Pr_2Zr_2O_{7-\delta}$ (Fig. 2(b)). This could be explained by oxygen vacancies in $Pr_2Zr_2O_{7-\delta}$ being in different Wykoff positions in comparison to those in $Yb_2Ti_2O_{7-\gamma}$. We will consider this further in the Discussion.

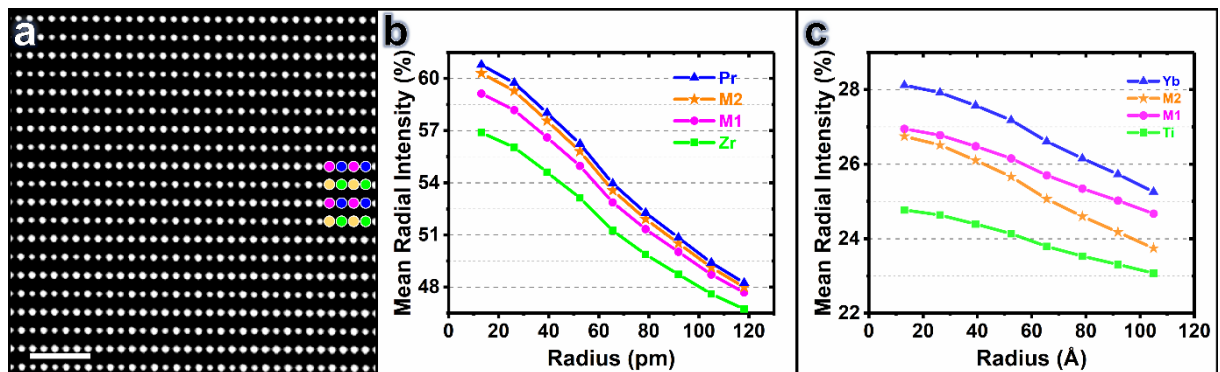

**Fig. 2.** (a) ADF-STEM image taken with an ADF detector inner angle of 2.4α from sample B ($Pr_2Zr_2O_{7-\delta}$) viewed along [2 1 1] (scale bar is 1 nm). The four visible atom columns in this image, i.e. Pr, Zr, *M1* and *M2* columns, are shown in blue, green, pink and orange at the top right of the image, respectively. (b) Mean radial intensity profiles for the four atom columns, i.e. Pr, Zr, *M1* and *M2*, in (a) calculated using DetectColumns [15] program. (c) Mean radial intensity profiles for Yb, Ti, *M1* and *M2* atom columns in the ADF image of pyrochlore $Yb_2Ti_2O_{7-\gamma}$ (Sample 2 in Ref. [15]) adapted from Ref. [15].

We also investigate the effect of oxygen atoms on the shape of their nearby cation columns in $LaFeO_3$ with a perovskite structure. Fig. 3(b) shows an ADF-STEM image viewed along [110] direction obtained from $LaFeO_3$. As schematically shown in Fig. 3(a), there are two types of cation columns in this projection; those containing only La (displayed in blue) and those containing only Fe (displayed in pink and yellow). The Fe columns can be divided into two different columns, *Fe1* and *Fe2*, according to the position of their nearby oxygen columns. As shown in Fig. 3(a), the oxygen atoms around *Fe1* columns are close to the vertical (*y*) axis of the image while those around *Fe2* columns are close to the horizontal (*x*) axis. As the number of oxygen atoms and their distance from the centre of cation columns are the same for both the *Fe1* and *Fe2* columns, the intensity of these atom columns in the ADF-STEM images should be similar. The intensities of atom columns shown in Fig. 3(b) are plotted as histograms in Fig. 3(c) and, as expected, there is no difference between the intensity of the *Fe1* and *Fe2* columns. Nevertheless, the width of the *Fe1* and *Fe2* columns are not the same. The calculated standard deviations obtained from two-dimensional (2D) Gaussian fitting on Fe columns (i.e. width of Fe columns) in ADF images obtained from LaFeO$_3$ samples are illustrated in Figs. 3(d) and 3(e). We observed that the *Fe2* columns are broader than the *Fe1* columns along the *x* direction (Fig. 3(d)) while the *Fe1* columns are broader along the *y* direction (Fig. 3(e)). These results are consistent with the position of oxygen atoms along *x* and *y* around these columns. In order to prove that the observed difference in the shape of the *Fe1* and *Fe2* columns is due to the difference in the position of their nearby oxygen atoms, we performed ADF-STEM image simulations using the Prismatic [44, 45] software. The simulation results (Fig. 3(f)), consistent with the experimental results (Fig. 3(d) and 3(e)), indicate the *Fe2* columns are broader along the *x* direction whereas the *Fe1* columns are broader along the *y* direction; i.e. the oxygen atoms change the shape of their nearby cation columns.

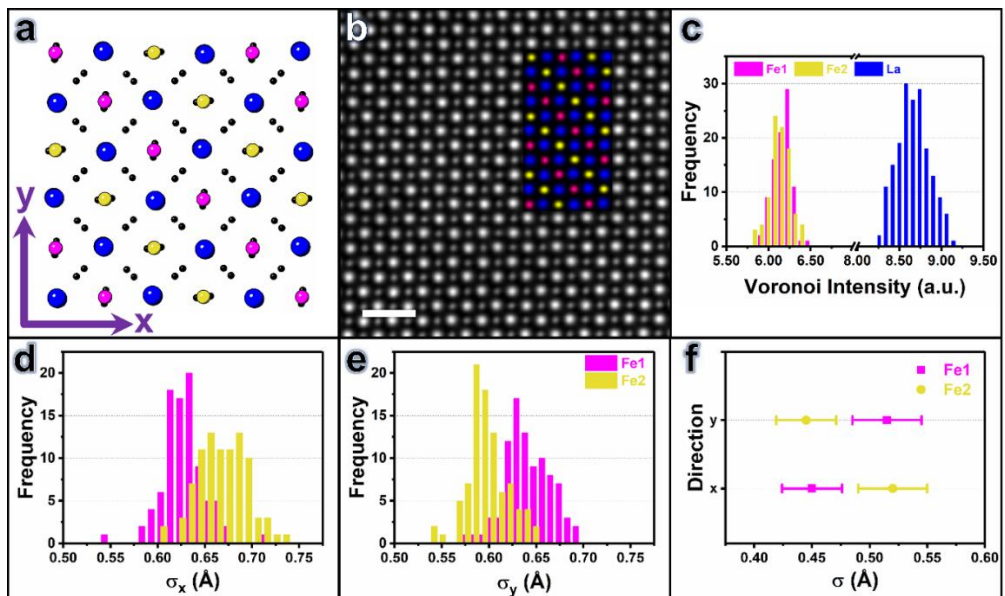

**Fig. 3.** (a) [1 1 0] projection of the crystal structure of LaFeO$_3$. Here, La and O atoms are displayed in blue and black, respectively. The Fe atoms, whose nature depends on the position of their nearby oxygen atoms, are shown in pink (the *Fe1* columns) and yellow (the *Fe2* columns). (b) ADF-STEM image taken with an ADF detector inner angle of 4.6α from the perovskite LaFeO$_3$ viewed along [1 1 0] (scale bar is 1 nm). The *2D* Gaussian fitting image on the three visible atom columns in this image is overlayed in which the La, *Fe1* and *Fe2* columns are shown in blue, pink and yellow, respectively. (c) Histograms of the extracted mean integrated intensities from (b) using Voronoi cells centred on each atomic column. (d) and (e) display the histograms of the standard deviations of the Fe columns along the horizontal (*x*) and vertical (*y*) axes obtained from the *2D* Gaussian fitting on those columns shown in (b), respectively. (f) Standard deviations of the simulated Fe columns along the *x* and *y* axes.

*3.2 Electron energy loss spectroscopy*

The change in colour of praseodymium zirconate from brown ($Pr_2Zr_2O_7$) to green ($Pr_2Zr_2O_{7-\delta}$) after reduction in Ar + 3%$H_2$ atmosphere is due to a reduction of the $Pr^{4+}$ cations to $Pr^{3+}$ [20, 47]. As electron energy loss near-edge structure (ELNES) spectra can be used to obtain information on the Pr valance state in praseodymium oxides [48, 49], we examine the oxidation state of Pr in our praseodymium zirconate samples using Pr-$M_{5,4}$ ELNES spectra. Pr shows two white lines $M_5$ and $M_4$, due to electron transitions from $3d_{5/2}$ and $3d_{3/2}$ subshells to the outer unoccupied $4f$ states, respectively. In praseodymium oxides, the number and position of peaks in the Pr-$M_{5,4}$ ELNES spectrum depends on the valence state of the Pr atoms. For instance, the $Pr^{3+}$-$M_4$ white line shows a shoulder on its low energy side that is absent in $Pr^{4+}$-$M_4$ [49]. In addition, a decrease in the Pr valence state leads to (a) a shift in the Pr-$M_{5,4}$ peaks to lower energies and (b) a decrease in the intensity ratio $I_{M_5}/I_{M_4}$ which is called the branching ratio [49].

The experimental Pr-$M_{5,4}$ ELNES spectra obtained from sample A and sample B are shown in Fig. 4. Peak energies, extracted by fitting the data to three Lorentzian curves, as well as intensity ratios are listed in Table 1. As shown in Fig. 4 and Table 1, the energy losses as well as the branching ratio for Pr-$M_{5,4}$ decrease from the brown sample to the green one. In other words, the Pr valence state reduced from $Pr_2Zr_2O_7$ to $Pr_2Zr_2O_{7-\delta}$, i.e. Pr–O bonds in praseodymium zirconate are affected by reduction in Ar + 3%$H_2$ atmosphere.

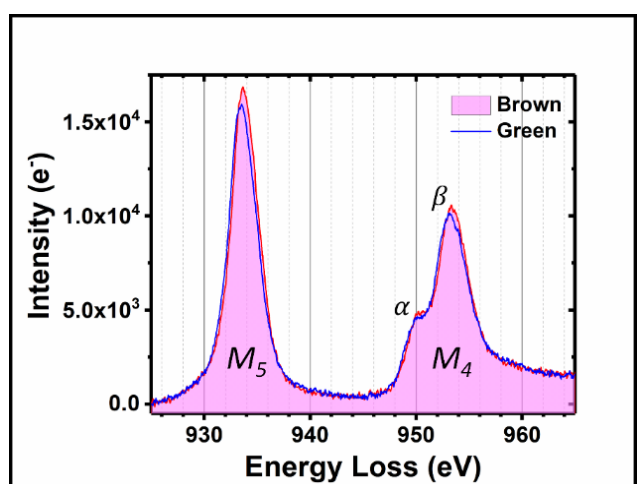

**Fig. 4.** Experimental EEL Pr-$M_{5,4}$ spectra of sample A and sample B ($t/\lambda$ = 0.35, where $t$ is the sample thickness and $\lambda$ is the inelastic mean free path of electron in the samples).

Table 1. Pr ELNES energies (eV) and branching ratios for the Pr pyrochlore samples.

| | | Sample A | Sample B |
|---|---|---|---|
| Pr | $E_{M_5}$ | 933.74 ± 0.01 | 933.56 ± 0.01 |
| | $E_{M_4}(\alpha)$ | 950.11 ± 0.03 | 949.83 ± 0.03 |
| | $E_{M_4}(\beta)$ | 953.37 ± 0.02 | 953.22 ± 0.02 |
| | $I_{M_5}/I_{M_4}$ | 1.30 ± 0.02 | 1.23 ± 0.02 |

## 4. Discussion

The ADF results obtained from the oxygen-deficient pyrochlores praseodymium zirconate and ytterbium titanate indicate that oxygen vacancies change the intensity distribution around *M1* and *M2* columns in both materials, but in different ways. According to the configuration of oxygen atoms around the *M* columns shown in Fig. 1(b), we expect a decrease in the occupancy of oxygen at *48f* sites to reduce the ADF intensity of both the *M1* and *M2* columns. However, as the distance between O-*48f* atoms and the centre of the *M2* columns (~50 pm) is much smaller than their distance from the centre of the *M1* columns (~100 pm), the reduction in the radial intensity of the *M2* columns caused by O-*48f* vacancies should mainly occur at small radii. Thus, the decrease in intensity for *M2* columns at small radii for $Yb_2Ti_2O_{7-\gamma}$, in comparison with stoichiometric $Yb_2Ti_2O_7$, can be attributed to O vacancies at *48f* sites. In a pyrochlore structure ($A_2B_2O_7$), O-*48f* are located in tetrahedral voids created by two *A*-site and two *B*-site atoms (Fig. 5(a)) while O-*8b* are surrounded by only *A*-site cations (Fig. 5(b)). This means that *B*-site cation reduction can only be caused by oxygen vacancies at *48f* sites. The EELS data presented in Ref. [15] showed that Ti exhibits a lower oxidation state in $Yb_2Ti_2O_{7-\gamma}$ (Sample 2 in Ref. [15]) compared to $Yb_2Ti_2O_7$ (Sample 1 in Ref. [15]) which means that the oxygen vacancies are at *48f* sites in $Yb_2Ti_2O_{7-\gamma}$. This is consistent with our radial intensity measurements.

Following the same reasoning, we expect a reduction in the ADF intensity only of *M1* columns for a decreased oxygen occupancy at *8b* sites. Thus, the lower radial intensity of the *M1* columns, even at large radii, compared to *M2* columns in the ADF image obtained from $Pr_2Zr_2O_{7-\delta}$ (Fig. 2) can be explained by the presence of oxygen vacancies at *8b* sites in this material. O-*8b* are located in tetrahedral voids created by four *A*-site atoms in a pyrochlore structure (Fig. 5(b)) and therefore O-*8b* vacancies only affect the oxidation state of *A*-site atoms. Our EELS results showed that Pr exhibits a lower oxidation state in $Pr_2Zr_2O_{7-\delta}$ compared to $Pr_2Zr_2O_7$, which means that the oxygen vacancies are at *8b* positions in $Pr_2Zr_2O_{7-}$ $_\delta$. This is consistent with our radial intensity measurements indicating that the oxygen vacancies in $Pr_2Zr_2O_{7-\delta}$ are at *8b* positions. Thus, the off-centring of $Pr^{3+}$ ions from their ideal crystallographic positions which has been demonstrated as the origin of the complex magnetic ground state of $Pr_2Zr_2O_7$ [20-23]) is due to Coulomb repulsion between the O-*8b* vacancies and the Pr cations. Our STEM technique allows us to observe the local defects in $Pr_2Zr_2O_{7-\delta}$ which were undetected in x-ray diffraction and neutron scattering measurements reported earlier [20-23].

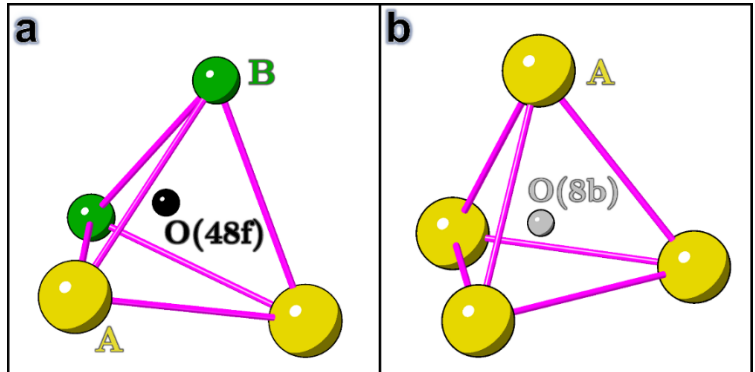

**Fig. 5.** Atom configuration around (a) O(*48f*) and (b) O(*8b*) in a pyrochlore $A_2B_2O_7$ structure.

The above results show that, while oxygen columns are invisible in ADF images, they make measurable contributions to the intensity of nearby cation columns. Careful comparisons of cation column intensity distribution for different samples can reveal changes in position, as well as occupancy, of oxygen atoms. The contribution of invisible light atoms to the contrast of ADF images can be direct or indirect, and we will return to this point later.

We can capture the essence of the behaviour in a very simple model by taking the two-dimensional ADF intensity distribution of a mixed atom column $I_V(x,y)$ to be the sum of scattering from a column of heavy cations $I_C(x,y)$ and adjacent light atoms $I_L(x,y)$, plus a second order term $I_{(C,L,\vec{d})}(x,y)$ due to the enhanced scattering that results from increased electron channelling when atom columns are very close to each other [50-52]. The latter depends on not only the type of atom in both columns but also their relative displacement ($\vec{d}$). So, the ADF intensity distribution of a visible cation column in an ADF image can be written as:

$$I_V = I_C + I_L + I_{(C,L,\vec{d})} . \qquad (2)$$

For example, the intensity of the *Fe1* and *Fe2* columns in $LaFeO_3$ can be written as:

$$I_1 = I_{Fe_1} + I_{O_1} + I_{(Fe_1, O_1, \vec{d}_1)}, \tag{3}$$

$$I_2 = I_{Fe_2} + I_{O_2} + I_{(Fe_2, O_2, \vec{d}_2)}. \tag{4}$$

In the [001] projection of Fig. 3 the number of iron and oxygen atoms, and the distance between them, is identical for both *Fe1* and *Fe2* columns in a specimen of fixed thickness. Taking bonding effects to be unimportant we may assume radially-symmetric scattering for $I_{Fe}$ which means the only difference in $I_1$ and $I_2$ is the relative displacement of the Fe and O atom columns in the second and third terms of each equation. If we subtract the intensity distribution of *Fe2* columns from that of *Fe1*, the remaining intensity is

$$I_1 - I_2 = I_{O_1} - I_{O_2} + I_{(Fe, O, \vec{d}_1)} - I_{(Fe, O, \vec{d}_2)}. \tag{5}$$

That is, scattering from Fe columns cancels, leaving only an intensity distribution that is dominated by the oxygen columns. Accordingly, for Fig. 3 we calculated the average intensity distribution of all *Fe1* columns (i.e. $\bar{I}_1$), all *Fe2* columns (i.e. $\bar{I}_2$) and their difference (i.e. $\bar{I}_1 - \bar{I}_2$) which are displayed in Figs. 6(a-c), respectively. Very careful examination of Figs. 3(a) and 3(b) reveals that both the *Fe1* and *Fe2* columns have an oval shape, with their long axes matching the position of nearby oxygen atoms as shown on the left of Fig. 6(a) and Fig. 6(b), i.e. vertical and horizontal respectively. In Fig. 6(c), according to Eq. (5), we expect positive values to correspond to $O_1$ and negative values to $O_2$. We may separate the two by taking only positive values (Fig. 6(d)) or negative values (Fig. 6(e)). Using this method, the oxygen columns in $Yb_2Ti_2O_7$ with the pyrochlore structure were also resolved (see Section II in Supplemental Material).

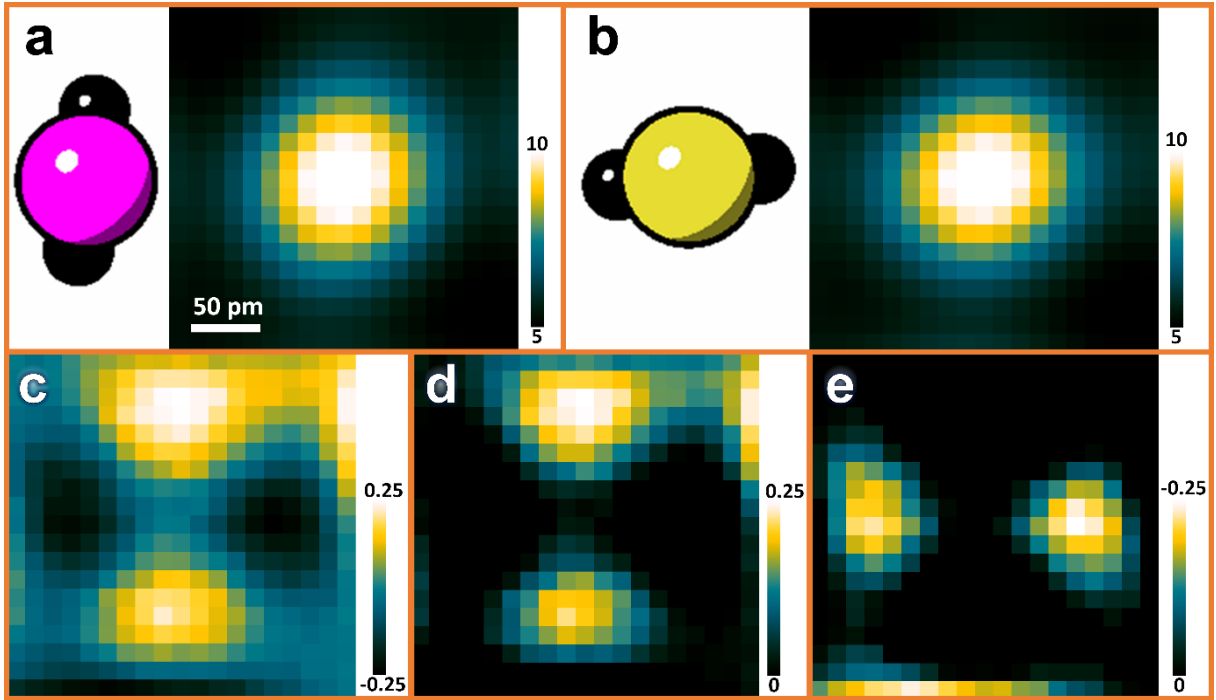

**Figure 6:** (a) Left: Schematic of the oxygen atom positions (black) around the Fe atoms (magenta) for the *Fe1* columns; Right: average ADF-STEM images obtained from all the *Fe1* columns shown in Fig. 3(b). (b) Left: Schematic of the oxygen atom positions (black) around the Fe atoms (yellow) for the *Fe2* columns; Right: average ADF-STEM images obtained from all the *Fe2* columns shown in Fig. 3(b). (c) Image of subtracted intensity for the average image of the *Fe2* columns from that of the *Fe1* columns, i.e. subtraction of (b) from (a). (d) Image of the positive values in (c) where the bright regions are consistent with the position of the oxygen atoms around the *Fe1* columns. (e) Image of the negative values in (c) where the bright regions are consistent with the position of the oxygen atoms around the *Fe2* columns.

The contrast of atom columns in ADF images is generated when a fraction of the incident electron beam is scattered to high angles by atoms in the specimen. As denoted in Eq. 1, the contribution of an oxygen column containing *m* oxygen atoms to the intensity of an ADF image would be proportional to $mZ^n$ ($Z_O = 8$) is significantly lower than that of a column containing heavy cations with the same number of atoms. In the case of LaFeO$_3$ with a thickness of ~40 nm (Fig. 3), regardless the effects of dynamic scattering such as the probe channelling and de-channelling, the contribution of an oxygen column to the intensity of the ADF image would be less than 2% and 7% of the contribution of a La and Fe column, respectively (where $m_{La} = m_{Fe} = 100$, $m_O = 50$ and $n = 1.7$). In reality, the contribution of oxygen atoms to the intensity of ADF image is much lower even than this, because the probe intensity ($J(r)$ in Eq. 1) at the position of oxygen atom columns is much lower than that at heavy cation atom columns. In fact, when the probe is at the oxygen column positions, the probe is affected both by de-channelling away from oxygen columns, and channelling along heavy cation columns [40, 41].

This effect is shown in the simulation of Fig. 7 for different incident probe positions. When the probe is at the oxygen column positions (probe positions: *P1* and *P3* in Fig. 7(b)) or close to them (e.g. *P4*), the probe intensity spreads considerably on, and oscillates around, their nearby iron column. This will be the case even if there is no oxygen column around the Fe column (see Section IV in Supplemental Material) due to the attractive Coulombic potential of the heavy iron atoms. However, as shown in Figs. 7(c-e) the beam intensity at the iron column is significantly higher in the presence of oxygen columns. For instance, as illustrated in Fig. 7(d), when the probe is at the *P3* position, the total beam intensity at the iron column in the presence of oxygen columns ($A_{Fe}$) is ~3.34 which is significantly higher (~27%) than that value in the absence of oxygen columns ($A_{Fe'} \approx 2.63$). Therefore, when the probe is at oxygen columns in the vicinity of heavy cation columns, the light oxygen atoms enhance spreading of the electron beam onto their nearby iron columns where heavy iron atoms, according to Eq. 1, can significantly contribute to the contrast of ADF images (see also Section V in Supplemental Material).

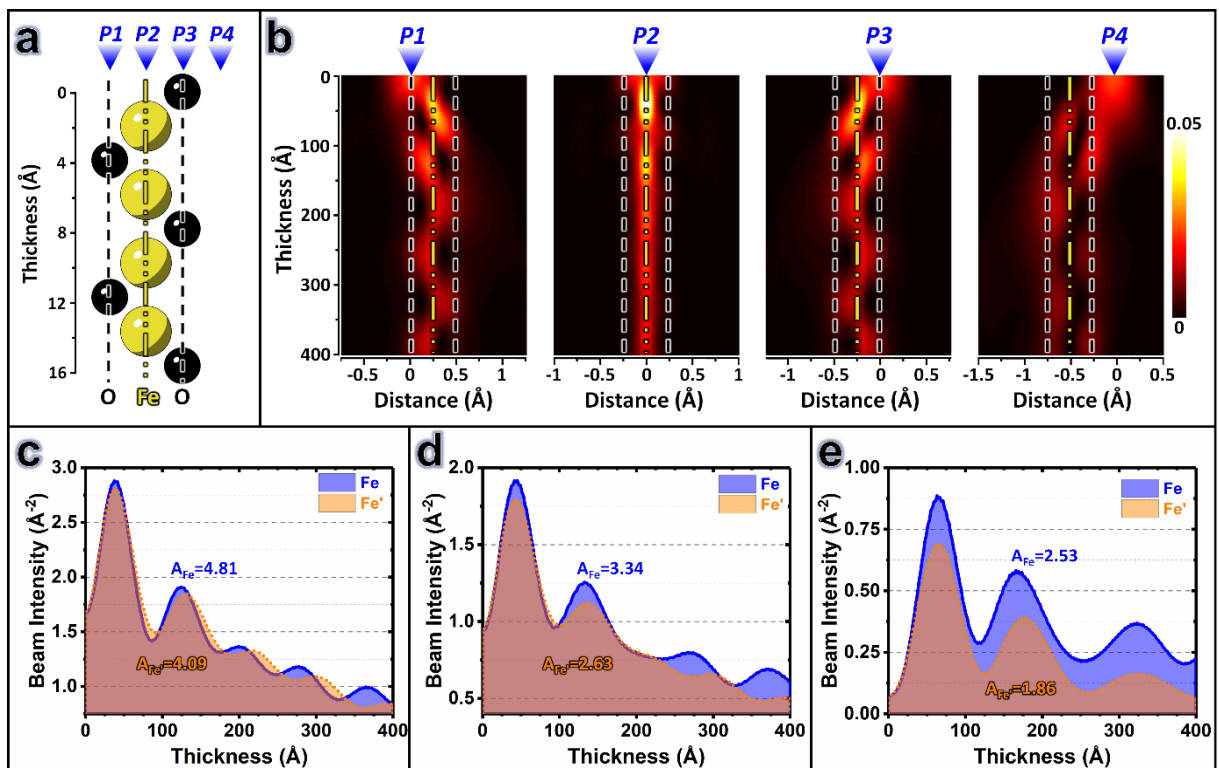

**Figure 7:** (a) Schematic of the atomic positions along an iron column and its nearby oxygen columns. (b) Simulated 2D beam intensity propagation profiles using Dr. Probe software [46] when the probe was at the *P1*, *P2*, *P3* and *P4* positions displayed in (a). Beam intensity depth profiles along the iron column in the presence of oxygen columns (blue lines) and in the absence of oxygen columns (orange dashed lines) when the probe is located at (c) *P2*, (d) *P3* and (e) *P4* positions. Here, $A_{Fe}$ and $A_{Fe'}$ show the integration of the beam intensity over 40 nm along iron columns in the presence of oxygen columns and in the absence of oxygen columns, respectively.

The contribution of oxygen atoms to a typical ADF image is thus so small that oxygen columns are effectively invisible. Only by subtraction of images, in which the heavy atom contribution cancels, can the contribution of oxygen columns be observed as shown in our experimental and simulation results (Figs. 1–3 and 6). The result of this procedure is critically dependent upon the noise level, whose intensity is uncorrelated in the two images being subtracted and thus remains in the difference image. In our experimental ADF-STEM data, a sum of many individual images, the mean signal to noise ratio is between 30:1 and 60:1, allowing the oxygen atom signal to easily be detected. We can also produce noisier data by summing fewer images, and find that while oxygen columns can also be traced in ADF images with lower SNRs (see Fig. S3 in Section III of Supplemental Material) a minimum SNR of roughly 10:1 is needed.

In summary, using the methods described here, it is possible to obtain (quantitative) information about light atoms (e.g. O) in materials containing both light and relatively heavy elements (e.g. oxides) using ADF-STEM provided that we use certain viewing directions (zone axes) for imaging in which the oxygen atoms are very close to atom columns containing heavy atoms. We expect the contrast of the resolved oxygen columns in the ADF images to contain quantitative information regarding the occupancy of oxygen sites. The methods can readily be used to investigate oxygen displacement/occupancy in other functional ceramic oxides with perovskite (e.g. $BaTiO_3$ [53, 54], $LaMnO_3$ [55], $Na_{0.5}Bi_{0.5}TiO_3$ [6], $BiFeO_3$ [56], etc.) or pyrochlore (e.g. $Dy_2Ti_2O_7$ [57], $Tb_2Hf_2O_7$ [58], etc.) structure where oxygen plays a key role in their physical and electrical properties. Furthermore, these methods can also be applied to characterize other invisible light atoms in the ADF images obtained from materials containing light and heavy elements, such as Li in lithium-ion battery materials [27, 59, 60].

## 5. Conclusions

We have investigated the atomic structure of three oxides with either a pyrochlore (praseodymium zirconate and ytterbium titanate) or a perovskite (lanthanum ferrite) structure using aberration corrected STEM. The intensity of visible atom columns in ADF images is sensitive to the number and position of their nearby invisible light atoms (i.e. oxygen). In addition, by examining the radial intensity profile of the cation columns, oxygen vacancies are observed at *8b* and *48f* Wykoff sites in oxygen-deficient praseodymium zirconate ($Pr_2Zr_2O_{7-\delta}$) and ytterbium titanate ($Yb_2Ti_2O_{7-\gamma}$), respectively. According to the atom configuration in pyrochlore $A_2B_2O_7$, oxygen vacancies at *8b* sites in $Pr_2Zr_2O_{7-\delta}$ or *48f* sites in $Yb_2Ti_2O_{7-\gamma}$ would be compensated by a reduction in the oxidation state of Pr or either Yb or Ti cations, respectively. EELS data is consistent with those radial intensity measurements and showed that Pr in $Pr_2Zr_2O_{7-\delta}$ and Ti in $Yb_2Ti_2O_{7-\gamma}$ have lower oxidation states in comparison to stoichiometric material. We also devised a simple technique by which invisible light atoms like O can be resolved in the ADF images. Using this technique we have, for the first time, resolved the invisible oxygen columns in ADF-STEM images obtained from perovskite and pyrochlore structures. We expect this approach to be applicable for any other material comprised of atoms with widely differing atomic number.

## 6. Acknowledgements

The authors acknowledge the Engineering and Physical Sciences Research Council (EPSRC) UK grants EP/N032233/1, EP/L017563/1 and EP/T005963/1.

# Supplemental Material

# Characterizing oxygen atoms in perovskite and pyrochlore oxides using ADF-STEM at a resolution of a few tens of picometers

Ali Mostaed[1], Brant Walkley[2,3], Monica Ciomaga Hatnean[4], Geetha Balakrishnan[4], Martin R. Lees[4], Richard Beanland[4], Derek C. Sinclair[2] and Ian M. Reaney[2]

[1] *Department of Materials, University of Oxford, Parks Road, Oxford OX1 3PH, UK*
[2] *Department of Materials Science and Engineering, University of Sheffield, Sheffield S1 3JD, UK*
[3] *Department of Chemical and Biological Engineering, University of Sheffield, Sheffield S1 3JD, UK*
[4] *Department of Physics, University of Warwick, Gibbet Hill Road, Coventry CV4 7AL, UK*

## 7. Crystallographic information used for image simulations

**Table S1**. Crystallographic parameters considered for a perfect LaFeO$_3$ crystal (*Pbnm* - ICSD 78062) with the lattice parameters of $a$ = 5.55702 Å, $b$ = 5.56521 Å and $c$ = 7.85426 Å.

| Atom | Fractional coordinates | | | Debye Waller factor (Å$^2$) |
|---|---|---|---|---|
| La | 0.9923 | 0.0292 | 0.25 | 0.52 |
| Fe | 0 | 0.5 | 0 | 0.73 |
| O1 | 0.0748 | 0.4855 | 0.25 | 0.37 |
| O2 | 0.7191 | 0.2817 | 0.0394 | 0.35 |

## 8. Resolving oxygen columns in experimental images

ADF images obtained from Yb$_2$Ti$_2$O$_7$ (Fig. S1(a)) were also examined by our proposed technique to see whether this technique can be used to resolve oxygen columns in the ADF images obtained from materials with the pyrochlore structure as well as the perovskite structure. As illustrated in Fig. S1(b), cation columns containing 50%Yb and 50%Ti with different positions of oxygen atoms around them are viewed in the [110] projection of Yb$_2$Ti$_2$O$_7$, i.e. those columns shown in cyan (named the *M3* columns) and orange (named the *M4* columns). Similar to the way we resolved oxygen columns in LaFeO$_3$ (Eqs. (1-4) and Fig. 6 in the main manuscript), we subtracted the average image of all the *M4* columns (i.e. $\bar{I}_{M4}$) from the average image of all *M3* columns (i.e. $\bar{I}_{M3}$) in Fig. S1(a) and displayed the positive and negative values of the subtracted image in Fig. S1(c) and Fig. S1(d), respectively. The bright contrast in Fig. S1(c) and Fig. S1(d) are consistent with the position of oxygen columns around *M3* and *M4*, respectively. In addition, it is possible to resolve oxygen columns around each individual *M3* or *M4* columns (i.e. $I_{O3_i}$ and $I_{O4_j}$) by filtering the negative values in the following equations (see Fig. S1(e) and Fig. S1(f)):

$$I_{O3_i} = I_{M3_i} - \bar{I}_{M4} \qquad i = 1,2,\ldots,n \tag{S1}$$

$$I_{O4_j} = I_{M4_j} - \bar{I}_{M3} \qquad j = 1,2,\ldots,m \tag{S2}$$

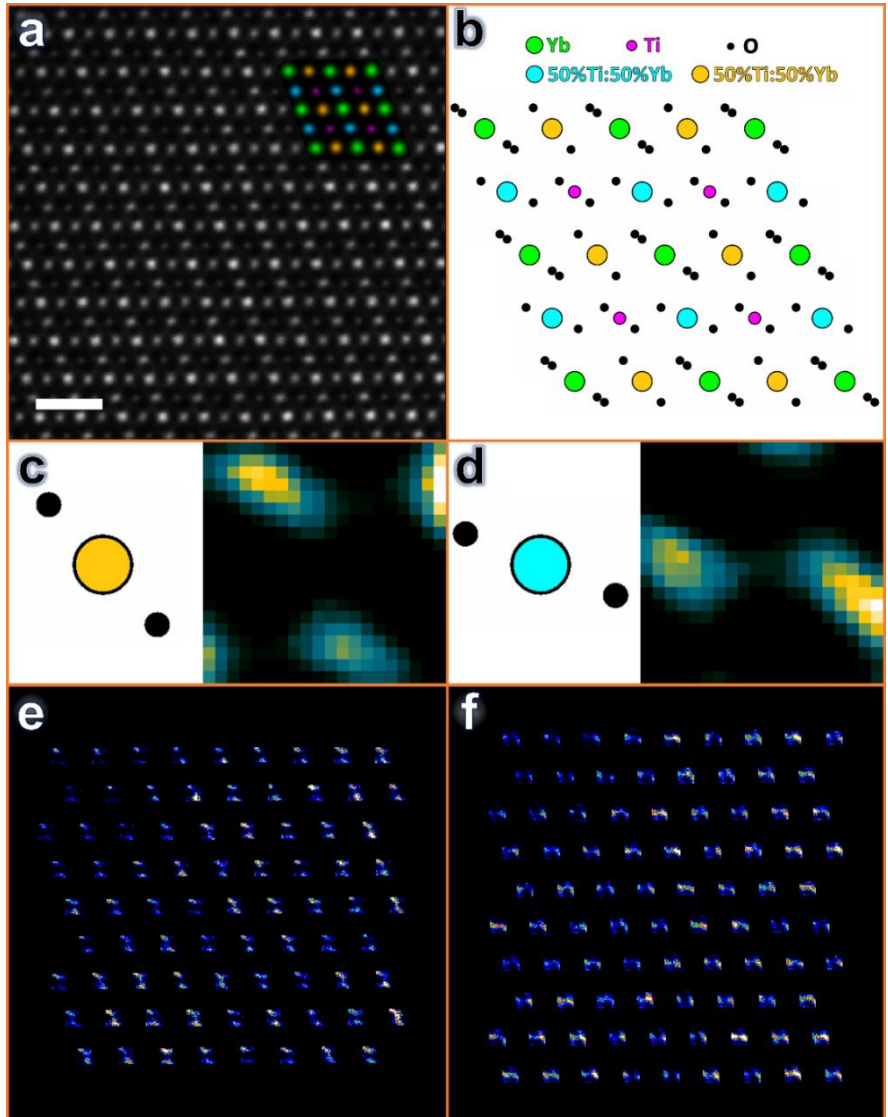

**Figure S1:** (a) ADF-STEM image taken with an ADF detector inner angle of 4.6α from the pyrochlore $Yb_2Ti_2O_7$ (Sample 1 in Ref. [1]) viewed along [1 1 0] (scale bar is 1 nm). (b) [1 1 0] projection of the crystal structure of $Yb_2Ti_2O_7$. Here, Yb, Ti and O columns are displayed in green, pink and black, respectively. Mixed cation columns (50%Yb:50%Ti) depends on the position of their nearby oxygen atoms are shown in cyan (named the *M3* columns) and orange (named the *M4* columns). (c) Left: Schematic of the oxygen atom positions around the mixed cation columns for the *M4* columns; Right: Image of the positive values in subtracted intensity from the average image of all the *M4* columns in (a) from that of the *M3* columns. (d) Left: Schematic of the oxygen atom positions around the mixed cation columns for the *M3* columns; Right: Image of the positive values in subtracted intensity from the average image of all the *M3* columns in (a) from that of the *M2* columns. (e) and (f) show the resolved oxygen columns around *M2* and *M1* columns in (a), respectively. Intensity calculations used in this figure were performed with the DetectColumns [1] program.

It is noteworthy that the above method to map oxygen columns around all individual cation columns in an ADF image is possible only if there is not a thickness variation across the sample in the imaging area.

Similarly we mapped oxygen columns around Fe columns in the ADF image obtained from LaFeO$_3$ (Figure S2(b)). In several cases, the resolved oxygen columns are not in the expected configurations (i.e. like Figs. 6(d) or 6(e)). This is likely related to the complex interplay between oxygen vacancies and octahedral tilting. Tilting occurs through a cogwheel like rotation pivoted at the corner shared apices of each octahedral [2]. In a perfect crystal, the magnitude of rotation or tilting is constant around the proscribed axes over long range. However, the presence oxygen vacancies inhibits the cogwheel rotation at the octahedral apices and locally reduces the magnitude of tilting [3]. Given sufficient oxygen vacancies, octahedral tilting is suppressed, or its onset driven to lower temperature. A natural consequence of the existence of oxygen vacancies in an octahedra-tilted framework therefore, is variation in the precise location of the oxygen atoms.

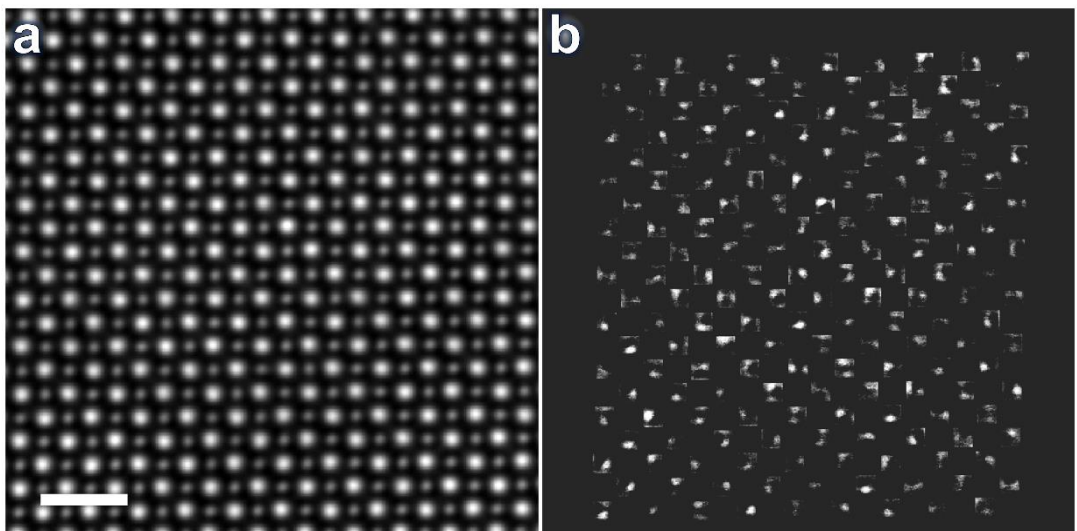

**Figure S2:** (a) ADF-STEM image taken from LaFeO$_3$ (see Fig. 3(b) in the main text) and (b) show the resolved oxygen columns around the iron columns in (a). Scale bar is 1 nm.

## 9. Signal to noise ratio (SNR) in an image

It is not possible to determine SNR from an individual experimental image, since the noise can only be estimated by comparison between the experimental image and a perfect, noise free image (which of course does not exist). However, the series of $N$ images we collect produces a mean image (sum/$N$) that has very low noise, which can be taken to be a good approximation of a perfect image and thus allows an accurate measure of SNR for individual images SNR(1). Furthermore, the SNR of the mean image, SNR($N$) can be obtained by extrapolation. The SNR for a pixel in individual images can be estimated as

$$SNR(1) = \frac{\bar{I}}{\sigma_1} \qquad (S1)$$

where $\bar{I}$ is the intensity of the pixel in the mean image from the whole series, and $\sigma_1$ is the standard deviation of the pixel's intensity across the series of individual images. Now, we may obtain $\sigma_2$ by making a new series, in which each image is the mean of adjacent pairs of the original images. This allows SNR(2) to be calculated. The process can be repeated, making image series that are comprised of the mean of triplets and quads, giving SNR(3), SNR(4), … . A plot of these values allows extrapolation to SNR(N).

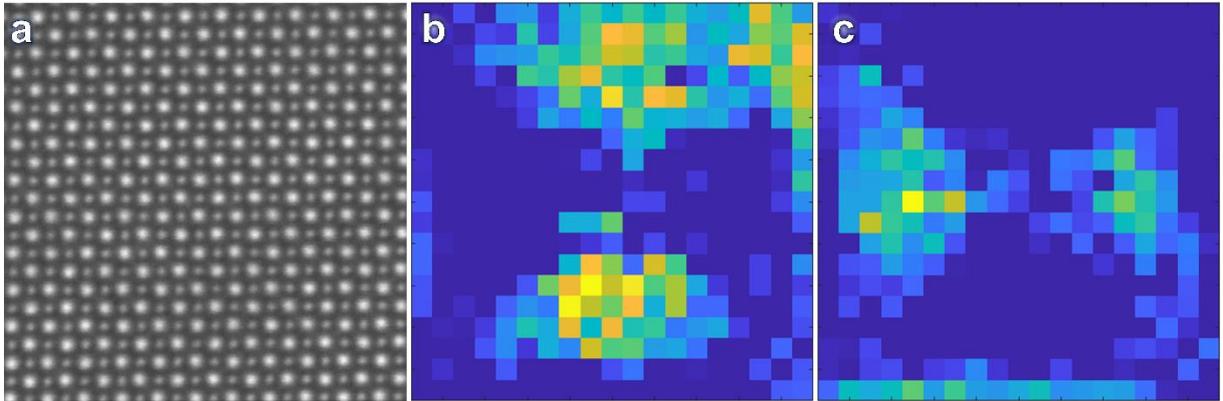

**Figure S3:** (a) ADF-STEM image taken from LaFeO$_3$ formed by summing three individual frames (SNR ~ 14). Position of the oxygen atoms around (b) the *Fe1* columns and (c) the *Fe2* columns measured using the similar method used for measuring Figs. 6(d) and 6(e).

## 10. Probe propagation close to an iron column

Similarly, our simulation results displayed in Fig. S3(b) indicate when the probe is close to an iron column (probe positions: *P1*, *P3* and *P4*), the probe intensity considerably spread on and oscillate around the iron column. This is due to the attractive Coulombic potential from the heavy iron atoms.

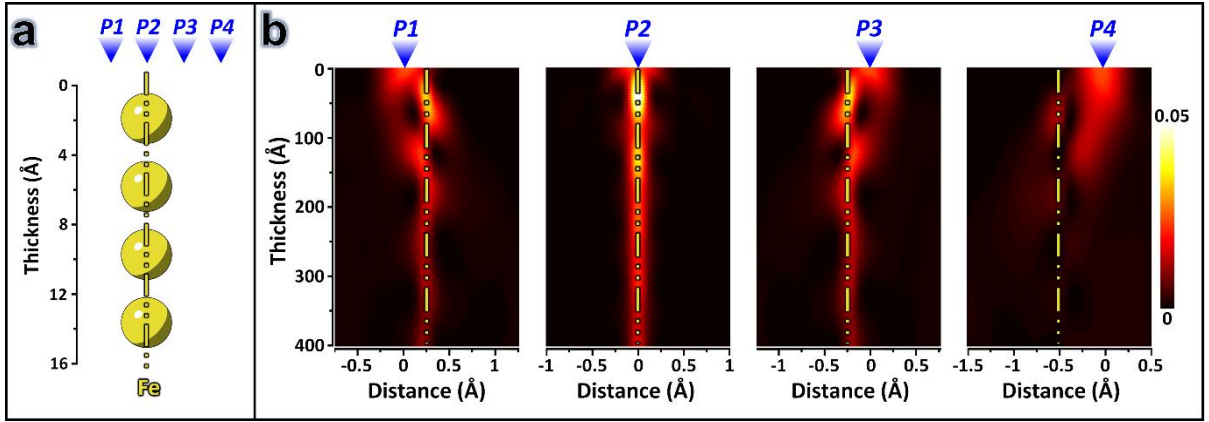

**Figure S4:** (a) Schematic of the atom positions along an iron column and the probe positions (*P1*, *P2*, *P3* and *P4*) along which the beam propagation has been simulated. (b) Simulated 2D beam intensity propagation profiles using Dr. Probe [4] software when the probe was at *P1*, *P2*, *P3* and *P4* positions displayed in (a).

## 11. Resolving oxygen column in simulated ADF image

As shown in Figs. S5(a-c), an oxygen column with a thickness of 30 nm can scatter only ~0.17% of the incident electron beam onto an ADF detector with collection angles between 70 to 280 mrad while a heavy cation column like an iron column scatters ~4.38% of the incident beam onto an ADF detector under the same conditions. Thus, the contrast of an oxygen column is less than 4% of an iron column and it is even lower ($< 2\%$) when compared to heavier atomic columns such as La and Pr. According to the data displayed in Figs. S5(a-c), the intensity of an iron column in the vicinity of an oxygen column ($I_{Fe-O}$) is about 10% higher than the intensity summation of a Fe column and O column ($I_{Fe} + I_O$). As we showed in Fig. 7, this increment in the intensity of the iron column is due to spreading electron beam onto the heavy iron atoms by oxygen atoms when the probe is located in the vicinity of the iron column. Furthermore, if we subtract the simulated ADF image obtained from an iron column (Fig. S5(d)) from the simulated ADF image obtained from an iron column with an oxygen column in its vicinity, i.e. *Fe-O* column, (Fig. S5(e)), we can clearly see that the areas with higher intensities in the *Fe-O* column are distributed around the position of the oxygen column. This means that the higher intensity of the *Fe-O* column is mainly due to the fact that the intensity of the oxygen column is intensified by the nearby iron column. Therefore, the intensity of the *Fe-O* column, according to Eq. 3, can be described as:

$$I_{Fe-O} = I_{Fe} + I_O + I_{(Fe,O,\vec{d})} \tag{S2}$$

where $I_{(Fe,O,\vec{d})} \approx 2.4 \times I_O$ in this case. The mechanism of this intensity intensification can be explained as follows. The electromagnetic field of oxygen atoms focused the beam along the oxygen column (electron channelling). Then, as this oxygen column is very close to the cation column containing Fe atoms, the focused electron beam by oxygen atoms can be scattered by their nearby Fe atoms which can scatter electrons to higher angles.

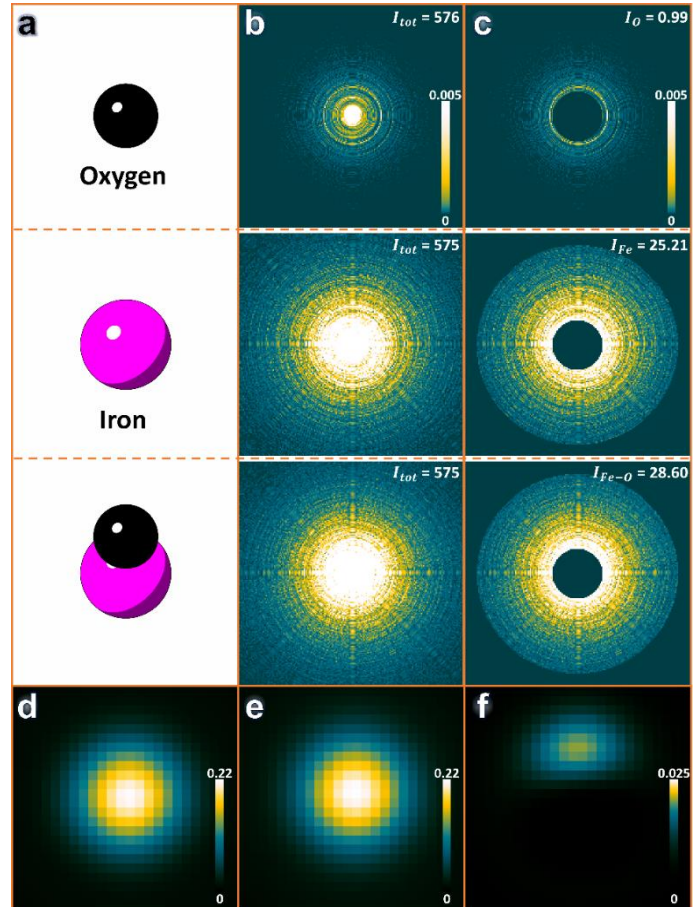

**Figure S5:** (a) Schematic of the atom positions for the O (top), Fe (middle) and Fe-O (bottom) columns used for image simulations. Note: the distance between O and Fe columns displayed in the bottom is 15 pm. (b) Simulated STEM diffraction for O, Fe and Fe-O columns over an area of 2.4 by 2.4 Å around the centre of each atom column. Here $I_{tot}$ is the total intensity of diffracted/scattered electrons by the atom columns in the STEM diffraction images. (c) Shows the fractions of the diffracted/scattered electrons displayed in (b) which can be collected by an ADF detector with the collection angle range of 70-280 mrad. (d) and (e) illustrate simulated ADF-STEM images obtained from the Fe and Fe-O columns, respectively. (f) Displays the image obtained from subtraction of (d) from (e). Image simulations used in this figure were performed with Prismatic [5, 6] software.